\documentclass[12pt,psfig,epsfig]{article}
\usepackage{graphicx}

\newcommand{\be}{\begin{equation}}
\newcommand{\ee}{\end{equation}}
\newcommand{\bea}{\begin{eqnarray}}
\newcommand{\eea}{\end{eqnarray}}

\begin{document}

\begin{center} {\Large \bf WHAT DOES QUANTUM CHAOS MEAN?}
\\ \vspace*{12 true pt} D.~Huard$^{a}$, H.~Kr\"{o}ger$^{a}$$\footnote{Corresponding author, Email: hkroger@phy.ulaval.ca}$,  
G.~Melkonyan$^{a}$, K.J.M.~ Moriarty$^{b}$ 
and L.P.~Nadeau$^{a}$
\vspace*{12 true pt}\\ $^{a}$ {\small\sl D\'{e}partement de Physique, Universit\'{e} Laval, Qu\'{e}bec, Qu\'{e}bec G1K 7P4, Canada} \\ 
$^{b}$ {\small\sl Department of Mathematics, Statistics 
and Computer Science, Dalhousie University, Halifax N.S. B3H 3J5, Canada}
\end{center}

\vspace*{12 true pt} \noindent {\bf Abstract}\\
We discuss how the concept of the quantum action can  be used to characterize quantum chaos. As an example we study quantum mechanics of the inverse square potential in order to test some questions related to quantum action. Quantum chaos is discussed for the 2-D hamiltonian system of harmonic oscillators with anharmonic coupling. 


\begin{center} {\bf 1. INTRODUCTION} \end{center}
 The classical chaos has no direct analogue in quantum physics. One reason is that in quantum mechanics (Q.M.) one cannot define a point in phase space (due to Heisenberg's uncertainty relation), hence one 
cannot define Lyapunov exponents from diverging trajectories, 
thus quantum chaos 
cannot be represented quantitatively via Poincar\'e sections.
In Ref.  (Cammetti et al., 2000) the effective action has been used to describe 
classically chaotic quantum systems. In Ref. (Jirari et al., 2001) the quantum action has been introduced.
It basically states that transition amplitudes in quantum mechanics can be expressed in terms of an action, which has the form of the classical action, but with parameters (mass, potential) which are different from those of the classical action. 
The quantum action can be interpreted as a renormalized action in Q.M. (Jirari et al., 2002).
It is similar to the standard effective action, 
however, it is free of the deseases of the latter (infinite series of higher derivative terms, non-localities etc.). 
In the limit of large imaginary transition time, the existence of the quantum action has been proven (Kr\"oger, 2002). 

The quantum action has been originally proposed as a conjecture obeying following requirements. For a given classical action 
\begin{equation}
S=\int d t \frac{m}{2}\dot{x}^2-V(x),
\end{equation}
there is a quantum action
\begin{equation}
\tilde{S}=\int d t \frac{\tilde{m}}{2}\dot{x}^2-\tilde{V}(x),
\end{equation}
which parametrizes the quantum mechanical transition amplitude (Green's function) $G(y,t,x,0)$ by the expression
\begin{equation}
G(y,t,x,0)=\tilde{Z}exp\left[-\tilde{\Sigma}(\tilde{x}_{cl})/\hbar\right],
\end{equation}
where $\tilde{x}_{cl}$ denotes the classical path corresponding to the action $\tilde{S}$ and 
$$
 \left.\tilde{ \Sigma}(\tilde{x}_{cl})=\tilde{S}[{x}_{cl}]\right|_{x,0}^{y,t},
$$
where $t$ is the transition time,  $x$ and $y$ are final and initial points of trajectory and $\tilde{Z}$ is normalization factor. 
A priori, it is not evident that such quantum action exists.
By now the existence of the quantum action has been established
in the following cases: (i) Harmonic oscillator. In this case the quantum action is identical to the classical action. (ii) In the limit when the transition time $t \to 0$, the quantum action exists. Dirac has noticed that the path integral in this limit is dominated by the contribution from the classical trajectory. Hence the quantum action coincides with the classical action. (iii) In imaginary time ($t\to -iT$ is necessary to describe thermodynamics at finite temperature) and going to the limit $T \to \infty$, the quantum action exists. In general it is quite different from the classical action.

\begin{figure}[tbh]
\vspace{9pt}
\begin{center}
\includegraphics[width=0.4\linewidth,angle=0]{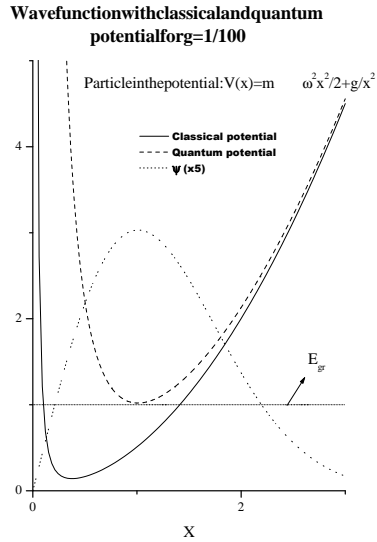}
\end{center}
\caption{Inverse square potential. Classical potential with  (full line), quantum potential (dashed line) and ground state wave function ( dotted line). }
\label{Fig_PotWave}
\end{figure}

This leads to the question: What about finite transition time? 
Does the quantum action exist for arbitray finite transition time? Does it parametrize Q.M. transition amplitudes for all $x$ to $y$ equally well? In general, the quantum action has to be determined non-perturbatively by numerical computations. What can be said about stability, convergence and errors of such procedure? What is quantum chaos in the framework of the quantum action? Those questions are the subject of the paper.
\begin{center} {\bf 2. INVERSE SQUARE POTENTIAL} \end{center}
We consider the following classical potential in 1-D
\be
\label{ClassPot}
V(x) = \frac{1}{2} m \omega^{2} ~ x^{2} + g ~ x^{-2} 
= v_{2} ~ x^{2} + v_{-2} ~ x^{-2} ~ .
\end{equation}
The corresponding classical action is given by
$S = \int dt ~ \frac{1}{2} m \dot{x}^{2} - V(x)$, while the 
classical action in imaginary time ($t \to -iT$) is given by 
$S_{E} = \int dt ~ \frac{1}{2} m \dot{x}^{2} + V(x)$ 
The potential is parity symmetric.
Because it has an infinite barrier (for $g > 0$) at the origin,
the system at $x < 0$ is separated from the system at $x > 0$. We consider only the motion in the domain $x > 0$. The potential is shown in Fig.[\ref{Fig_PotWave}]. We have chosen to consider the inverse square potential, because of the distinct feature that the corresponding quantum mechanical transition amplitudes are known analytically (Khandehar \& Lavande, 1975; Schulman, 1981, sec. 32).

 The Euclidean transition amplitude reads
\be
G_{E}(x,T;y,0)=\frac{m\omega \sqrt{xy} }{ \hbar \sinh(\omega T) } 
\exp \left\{- \frac{ m \omega} {2 \hbar} (x^{2} + y^{2}) \coth(\omega T) \right\}
I_{\gamma} \left( \frac{m \omega xy} {\hbar \sinh(\omega T) } \right) ~ ,
\end{equation}
where $I_{\gamma}$ is the modified Bessel function and $\gamma =  \left( 1 + 8 m g/\hbar^{2} \right)^{1/2}/2$. 
%
The transition amplitude contains all information on the spectrum and wave functions. For example, by going to the limit $T \to \infty$ (Feynman-Kac limit)
one reads off the ground state energy and wave function
\begin{equation}
E_{gr}= \hbar \omega (1 + \gamma), \quad \quad \psi_{gr}(x) =Z_{0}^{1/2} ~ x^{1/2 + \gamma} ~
\exp[-\frac{m \omega}{2 \hbar} x^{2} ] ~ .
\end{equation}
This is, of course, identical with the direct solution from the Schr\"odinger equation. The wave function is shown in Fig.[\ref{Fig_PotWave}].

{\noindent \bf 2.1. Dynamical time and length scales} \vspace*{12 true pt} \\ 
In this work we do a numerical study of the inverse square potentialal to test the quantum action. The numerical studies presented below have been done in imaginary time.
Our choice of the model has been influenced by the analytical solvability and less by the question if it plays a role in nature.
Consequently, absolute values in physical units of the model parameters (mass, potential parameters) are not of primary importance. We have expressed those parameters in dimensionless units. 
However, what is important are time and length scales, 
which are dynamically generated by the model, say a time scale $T_{sc}$ and a length scale $\Lambda_{sc}$. Those scales serve as reference values
to give a sense to statements like "for large transition times $T$", or a "small spatial resolution $\Delta x$", which means $T/T_{sc} >> 1$ and 
$\Delta x/\Lambda_{sc} << 1$, respectively. How to choose those dynamical scales? We have used as time scale $T_{sc} =1/E_{gr}$.

As a length scale one may introduce the analogue of the Bohr radius of the ground state wave function. Another possibility is to define a length scale $\Lambda_{sc}$ by
\be
\int_{0}^{\Lambda_{sc}} dx ~ |\psi_{gr}(x)|^{2} = 0.95 ~ ,
\end{equation}
i.e. the length which covers $95\%$ of the probability of the ground state wave function. For the classical action considered in this paper those scale parameters are $T_{sc} = 0.4$ and $\Lambda_{sc} \approx 2.35$.
%

{\noindent \bf 2.2.  Analytical and numerical results for the quantum action in the asymptotic regime} \vspace*{12 true pt} \\
In Ref.(Jirari et al., 2002) we have shown for the Euclidean asymptotic regime $T \to \infty$, that the following analytic relations exist between ground state energy $E_{gr}$, ground state wave function $\psi_{gr}$ and the quantum action.$ \tilde{V}_{min} = E_{gr}$, i.e. the minimum of the quantum potential gives the ground state energy.  
Next, the ground state wave function can be expressed in terms of the parameters of the quantum action 
\be
\label{GroundStateWave}
\psi_{gr}(x) = \frac{1}{N} ~ e^{ - \int_{\tilde{x}_{min}}^{x} dx' ~ 
\sqrt{2 \tilde{m}( \tilde{V}(x') - \tilde{V}_{min} ) }/\hbar } ~ .
\end{equation}
Finally, there is a relation between classical and quantum action (transformation law) which reads (for $x > x_{min}$)
\be
\label{TransformLaw}
2 m(V(x) - E_{gr}) =  
2 \tilde{m}(\tilde{V}(x) - \tilde{V}_{min}) 
- \frac{\hbar}{2} \frac{ \frac{d}{dx} 2 \tilde{m} (\tilde{V}(x) - \tilde{V}_{min})}
{ \sqrt{2 \tilde{m}( \tilde{V}(x) - \tilde{V}_{min} ) } } ~
\mbox{sgn}(x-\tilde{x}_{min}) ~ .
\end{equation}
Here $\tilde{x}_{min}$ denotes the position of the minimum of the quantum potential.
Let us make an ansatz for the quantum action, characterized by a mass $\tilde{m}$ 
and a quantum potential of the form
\be
\label{QuantPot}
\tilde{V}(x) = \tilde{v}_{2} x^{2} + \tilde{v}_{-2} x^{-2} + \tilde{v}_{0} ~ ,
\end{equation}
and see if it satisfies transformation law, Eq.(\ref{TransformLaw}).
Using Eqs.(\ref{ClassPot},\ref{QuantPot}), inserting it into Eq.(\ref{TransformLaw}), and comparing the coefficients of the terms $x^{2}$, $x^{-2}$ and $x^{0}$, one obtains
\begin{eqnarray}
\label{Asymptotic}
&& \tilde{m} ~ \tilde{v}_{2} = \frac{1}{2} m^{2} ~ \omega^{2}, \quad \tilde{m} ~ \tilde{v}_{-2} = \frac{1}{2} \hbar^{2} ~ [\frac{1}{2} + \gamma]^{2}, \quad E_{gr} = \hbar ~ [1 + \gamma] ~ .
\end{eqnarray}
For the case when the classical potential is given by the parameters $m=1$, $\hbar=1$, $\omega=1$ and $g=1$, this yields $\tilde{m} ~ \tilde{v}_{2} \to 0.5 \quad$ and $ \tilde{m} ~ \tilde{v}_{-2} \to 2$. 
The numerical solution, shown in Fig.[2]
gives $\tilde{m} ~ \tilde{v}_{2} \to 0.4999$ and $ \tilde{m} ~ \tilde{v}_{-2} \to 2.008$. 
We observe that analytical and numerical results agree well.
\begin{figure}[thb]
\begin{center}
\includegraphics[height=0.43\linewidth,angle=270]{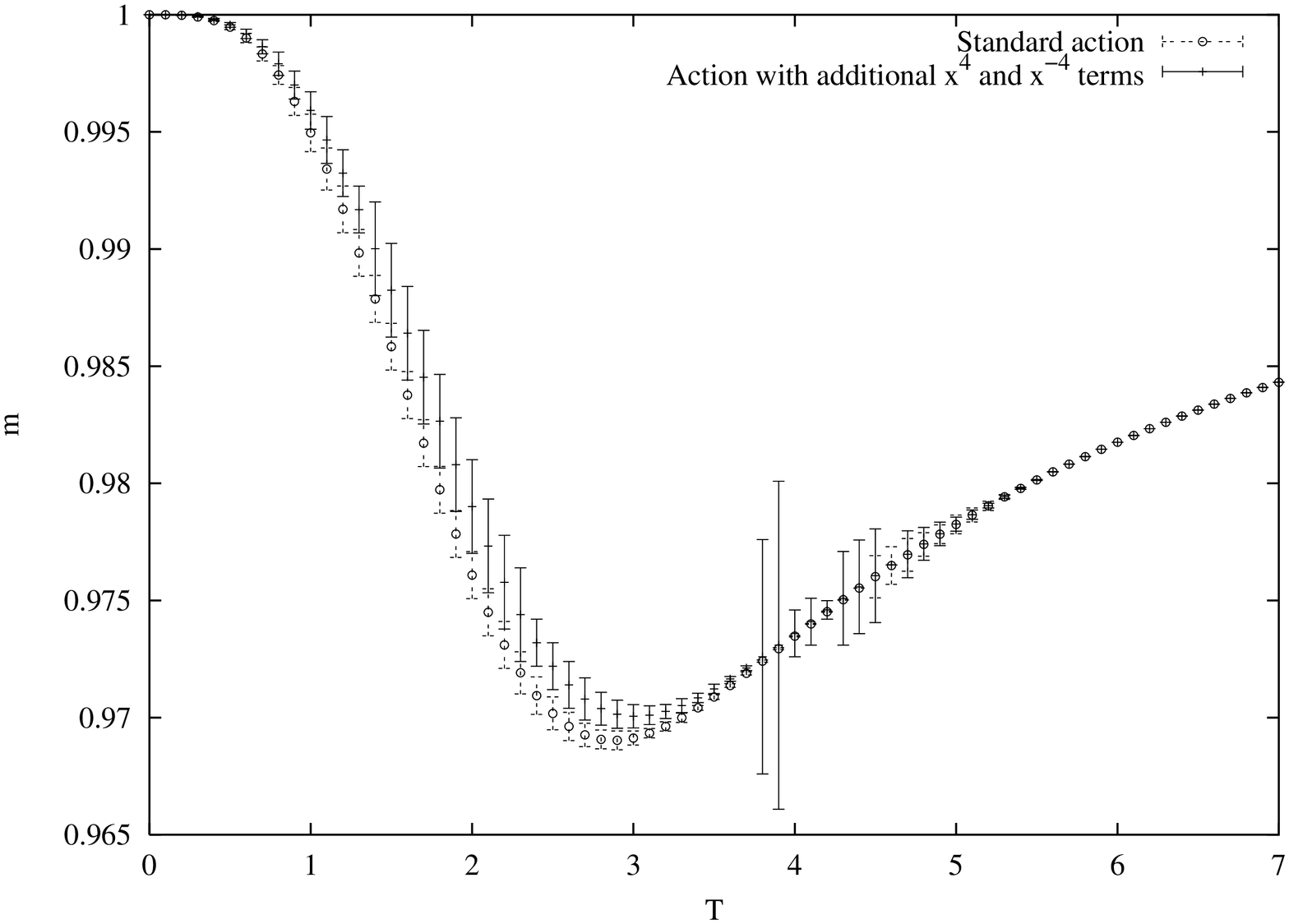}
\includegraphics[height=0.43\linewidth,angle=270]{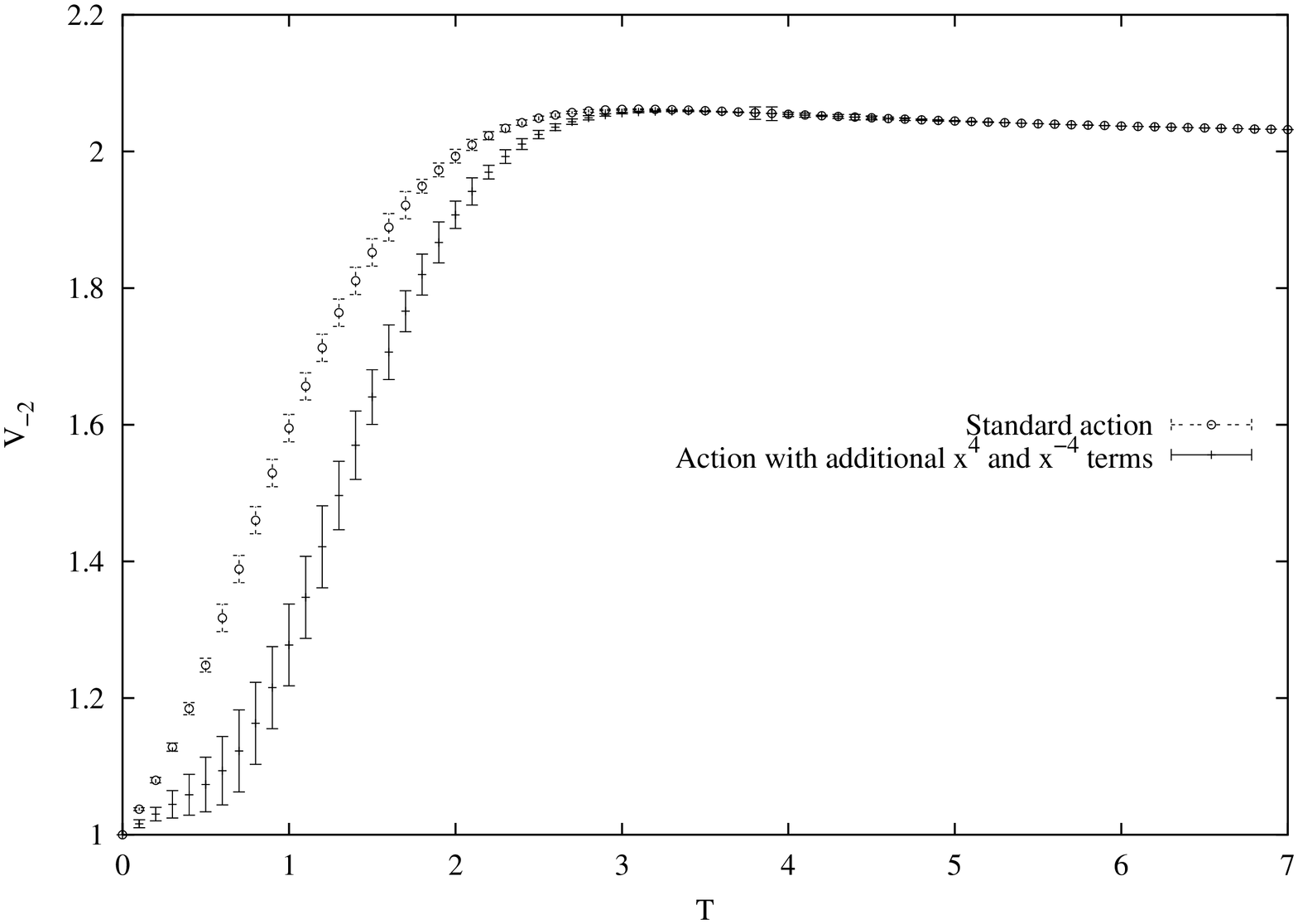}
\includegraphics[height=0.43\linewidth,angle=270]{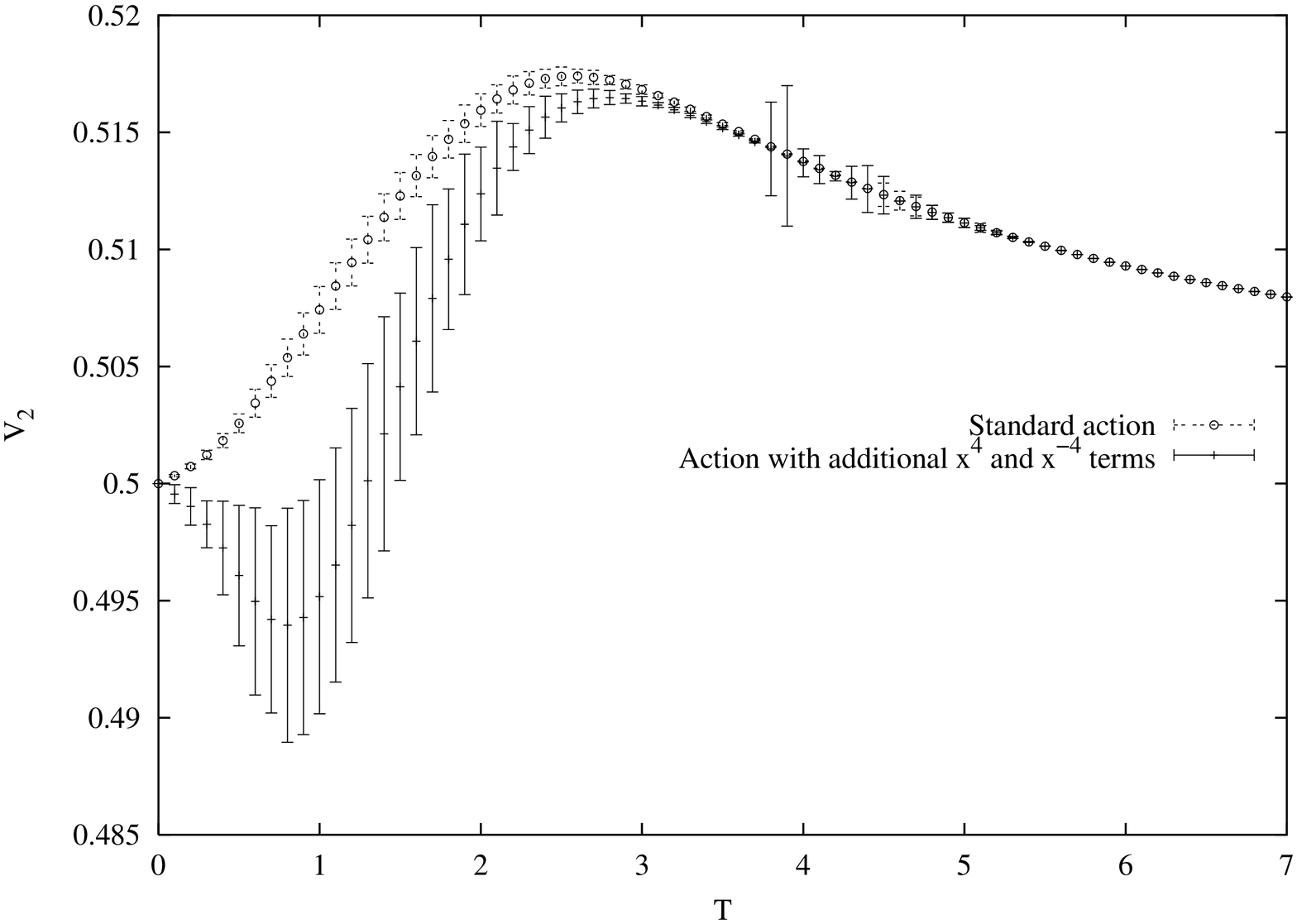}
\includegraphics[height=0.43\linewidth,angle=270]{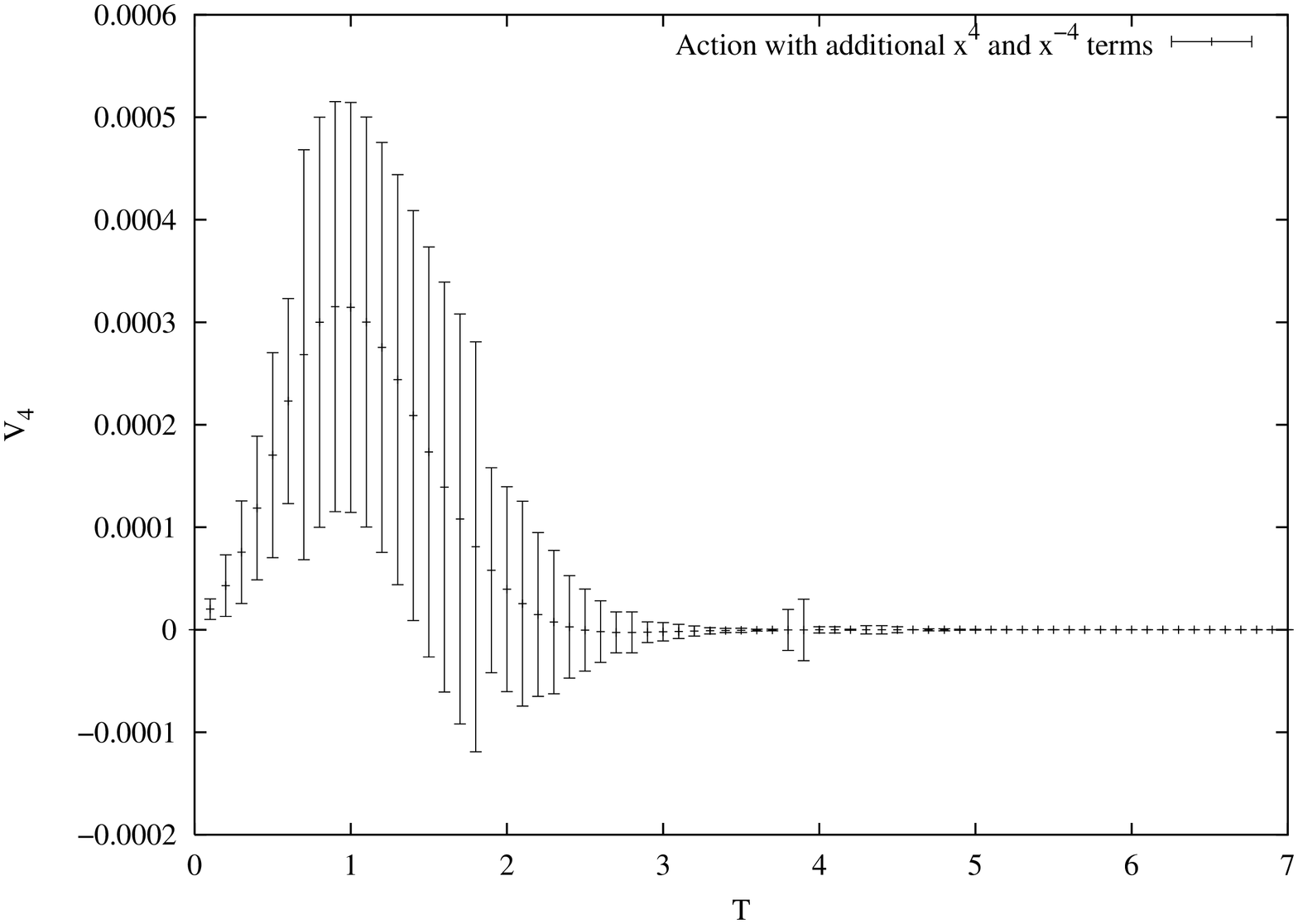}
\end{center}
\caption{Inverse square potential. Classical parameters $m=1$, $v_{2}=0.5$ ($\omega=1$), $v_{-2}=1$ ($g=1$). Boundary points of transition: $x_{i}$ - 2 points in interval $[4,5]$; 
$x_{f}$ - 10 points in interval $[0.5,3]$ and 
$\Delta t = 2 ~ 10^{-3}$. }
\label{Fig_ActPar}
\end{figure}

{\it Dependence on boundary points}. Next we look at the dependence of the quantum action parameters on the boundary points used in the fit. We have studied this in two ways:
(a) We kept the initial points $x_{i}$ fixed and varied the final points $x_{f}$.
(b) We kept the final points $x_{f}$ fixed and varied the initial points $x_{i}$.
In the first case, we have considered one initial point $x_{i}=0.3$ (kept fixed). 
As final points $x_{f}$ we took 100 points uniformly distributed in an interval and varied those intervals $[2,3]$, $[5,6]$, $[9,10]$ and $[2,10]$. 
 In the regime 
$T > 2 ~ (T > 5 T_{sc})$  there is no dependence on the location of the final boundary points. However, for $T < 2 ~ (T < 5 T_{sc})$ there is a noticeable dependence.

In the second case, as final boundary points $x_{f}$ we took 100 points uniformly distributed in the interval $[2,3]$ (interval kept fixed).
As initial points $x_{i}$ we took one point and varied it between 0.1 to 0.5.
Those results are qualitatively the same as in the first case.

One may ask: What is the reason for such dependence on boundary points? In our opinion the most likely explanation is that the quantum action is not an exact parametrisation of the transition amplitude in this regime. Hence the parametrisation depends on the parameters of the fit.

The two previous cases represent a strong imbalance between the number of initial versus final boundary points. As a third case we have considered a more balanced selection of bondary points. We took 10 initial points, uniformly distributed in the interval $[1.5, 2.5]$ and 10 final points, uniformly distributed in the interval $[1.1, 2.1]$.
The global maximum relative error in fitting the transition matrix elements is about $0.001$ in the regime $1<T<2$.

{\it Dependence on temporal resolution}. 
The determination of the parameters of the quantum action by a global best fit requires to solve the equation of motion from $\tilde{S}$ for all pairs of boundary points and to compute the value $\tilde{\Sigma}$ of the quantum action along such trajectory. Such numerical calculations depend on the temporal resolution $\Delta t$. We have studied the convergence of the parameters of the quantum action as a function of the density of meshpoints.
We parametrize the density of meshpoints by taking $N_{t}$ meshpoints per unit time interval $\Delta T=1$. We have varied $N_{t}$ from 200 up to 8000.
 For larger $T$, the results diverge. If one desires convergence for, say $T=8 \approx 20 T_{sc}$, one needs to double the density of meshpoints. Such behavior persists if we want to maintain convergence for even larger $T$. E.g., for $T=14 \approx 35 T_{sc}$, we need a 20-fold higher density of meshpoints ($N_{t}=8000$).
Such rapid increase in the density of meshpoints could signal chaos caused by  imperfect numerical solutions, finite internal precision, rounding errors etc.

 
\begin{center} {\bf 3. WHAT MEANS QUANTUM CHAOS?} \end{center}
Many systems in nature exhibit chaos (i.e. sensitivity and exponential dependence on initial conditions), but their quantum analogue  does not. The phenomenon of chaos is well understood and characterized in classical physics in terms of Poincar\'e sections and Lyapunov exponents. A system is said to be chaotic if it has positive Lyapunov exponents or equivalently their Poincar\'e section contains randomly distributed points.  

In general, there is no direct relation between classical and quantum dynamics ~( Thommen et al., 2003). Some interesting behaviour is seen in systems in the semiclassical limit. For
example, periodic orbits  in the classical systems are very important in
organizing the quantum eigenvalues and eigenfunctions.  Contrary to this when we are far from the semiclassical limit the phase space notion of classical mechanics cannot be used in quantum physics.  Heisenberg's uncertainty relation $\Delta q \Delta p> \hbar/2$ does not allow to define simultanously a single point in the phase space of particle motion. Even though the generalized basis is free from this difficulty the sensitivity and exponential dependence on initial conditions is absent due to  linearity of the Schr\"odinger equation. Necessary condition to have chaotic behavior  in classical physics is the nonlinearity introduced by dynamical variables $(q,p)$. In quantum physics the dynamical variable corresponds to the wave function, whereas the potential does not depend on the dynamical variable.  

The quantum action is free from these difficulties. First the quantum action  presents quantum mechanical 
 transition amplitude by a set of parameters $\tilde{m}$ and $\tilde{V}$. The quantum action has the same stracture as classical action, consequently there is no difficulty to define the phase space as in classical mechanics. Interestingly the quantum potential at large time limit ($T\to \infty$) is presented by the ground state wave function, which means the dynamics in the quantum potential is defined by the dynamical variable of the Sch\"rodinger equation.

Let us consider as an  example  the following classical system ~(Dahlquist \& Russberg, 1990), 
\begin{equation}
S=\int_0^T dt \frac{m}{2}\left( \dot{x}^2+\dot{y}^2\right)-V(x,y),
\end{equation}
with the potential 
\begin{equation}
V(x,y)=v_2(x^2+y^2)+v_{22}x^{2}y^{2},\quad \mbox{with} \qquad m=1,\quad v_2=0.5,\quad v_{22}=0.05.
\end{equation}
 This system is globally chaotic having small islands of regularity. To find the corresponding quantum action  we have parametrezed the quantum potential by analysing  the symmetries of the original Schr\"odinger equation. The time--reversal symmetry, parity conservation and symmetry exchange $x\leftrightarrow y$ allows one to parametrize the quantum potential as
\begin{equation}
\tilde{V}(x,y)=\tilde{v}_2(x^2+y^2)+\tilde{v}_{22}x^{2}y^{2}+\tilde{v}_{4}\left(x^{4}+y^{4}\right),
\end{equation} 
 where $\tilde{v}_2=0.504$, $\tilde{v}_{22}=0.05$ and $\tilde{v}_4\sim 10^{-5}$.

\begin{figure}
\begin{center}
\includegraphics[width=0.41\linewidth]{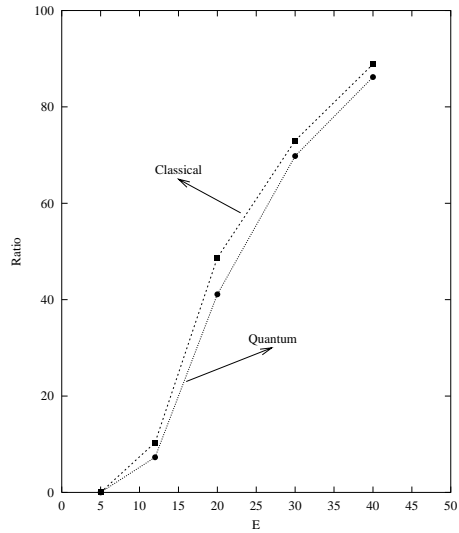}
\end{center}
\caption{The estimate of the chaotic volume of the phase space. $R$ is the ratio of the numeber of initial conditions with positive Lyapunov exponents to the total number of initial conditions. Anharmonic term $v_{22}=0.05$\label{FigR} }
\end{figure}

Numerical studies~(Caron et al.,2003) have shown the following behavior: For small $v_{22}$, Poincar\'e sections of classical and quantum systems are similar. With increase of energy differences between classical and quantum phase space become prononouced. This can be seen in Fig.~[\ref{FigR}], where the ratio of the chaotic phase space volume to total  phase space volume is presented for different values of energy.
From the discussed example follows that one can define the quantum chaos as~(Kr\"oger, 2002):
{\it Consider a classical system with action $S$. The corresponding quantum system displays quantum chaos, if the quantum action $\tilde{S}$ in the large time limit $T\to \infty$ generates a chaotic phase space}.

\begin{center} {\bf 4. CONCLUDING REMARK } \end{center}
We have presented the quantum action as a useful tool to study chaotic phenomenon in quantum systems. The structure of the quantum action allows to introduce the important notion of the phase space as in classical mechanics for quantum systems. 

\begin{center} { \bf ACKNOWLEDGEMENT } \end{center}
H.K. and K.M. are grateful for support by NSERC Canada. G.M. and D.H. have been supported in part by FCAR Qu\'ebec. 
For discussions and constructive suggestions H.K. is very grateful to A. Okopinska and L.S. Schulman.


\begin{center}{\bf REFERENCES}\end{center}
\begin{enumerate}
\item 
Cametti F., Jona-Lasinio G., Presilla C., \& Toninelli F., 
Proc. Int. Sch. of Phys. "Enrico Fermi", eds. G. Casati et al., IOS Press, Amsterdam (2000), p. 431.

\item
 Caron L.A., D.Huard, Kr\"oger H., Melkonyan G.,  Moriarty K.J.M., \& Nadeau  L.P. , quant-ph/0302133.

\item
 Dahlquist P. and Russberg  G., {\it Phys. Rev. Lett.} 66(1990)2837.

\item Jirari H. , Kr\"oger H.,  Luo X.Q., Moriarty  K.J.M.,  Rubin S.G.,  
{\it Phys. Rev. Lett.} 86(2001)187. 
 
\item
 Jirari H., Kr\"oger H., Luo  X.Q.,  Melkonyan G., Moriarty  K.J.M., 
{\it Phys. Lett. A} 303(2002)299. 

\item Khandekar K.C. \&  Lavande S.V., 
{\it J. Math. Phys.} 16(1975)384.

\item Kr\"{o}ger H.,
{\it Phys. Rev. A} 65(2002)052118.

\item
 Kr\"oger H., quant-ph/0212093.

\item Schulman  L.S., {\sl Techniques and Applications of Path Integration}, John Wiley \& Sons, New York (1981).

\item  Thommen Quentin, Garreau  Jean Claude, \& Zehnl\'e  V\'eronique  quant-ph/0306172.
\end{enumerate}
\end{document}